\documentclass[12pt]{article}
\usepackage{axodraw}
\usepackage[cp1251]{inputenc}
\usepackage[english]{babel}

\begin{document}

\title{
{\bf Intercepts of meson Regge trajectories in SU($N_c$) quantum chromodynamics 
with massless quarks}}
\author{A.A. Godizov\thanks{anton.godizov@gmail.com}\\
{\small {\it Institute for High Energy Physics, 142281 Protvino, Russia}}}
\date{}
\maketitle

\vskip-1.0cm

\begin{abstract}
By means of solving the Bethe-Salpeter equation with the minimal perturbative 
kernel (ladder approximation) in the arbitrary covariant gauge, there was found a 
series of intercepts of meson Regge trajectories associated with colorless 
singularities of the four-quark Green function in SU($N_c$) quantum chromodynamics 
with massless quarks. The gauge and renorm-invariance of the result is grounded.
\end{abstract}

\vspace*{1cm}

\section*{Introduction}

One of the most important and interesting problems of the theory of strong interaction is the problem of 
calculation of Regge trajectories (singularities of scattering amplitudes continued analytically 
to the region of complex angular momentum \cite{collins}) in the framework of quantum chromodynamics (QCD). 
Upon having the analytical behavior of Regge trajectories determined we automatically get 
information on hadron masses and decay widths corresponding to those points of Regge trajectories where 
trajectories themselves take on integer non-negative values. Notably, mass $M$ and decay width $\Gamma$ of 
some spin-$N$ hadronic state placed on Regge trajectory $\alpha(t)$ obey the equation 
$$
\alpha(t_N)=N\;\;\;({\rm Re}\,t_N=M^2,\;\;\;{\rm Im}\,t_N=-M\,\Gamma)\,.
$$ 
Moreover, the behavior of leading Regge trajectories in the region of small negative values of the argument 
governs the high energy evolution of the diffractive pattern at scattering to small angles. 

From the very nature of Regge trajectories in quantum field theory (poles 
of scattering amplitudes and corresponding Green functions) there follows one of their most important 
properties –- renorm-invariance, i.e., their independence of the renormalization scheme and the renormalization 
scale.

The most popular approach to calculation of Regge trajectories in QCD is the Balitsky-Fadin-Kuraev-Lipatov (BFKL) 
approach, which, in fact, brings us to looking for poles of solution of the 
linear integral BFKL equation \cite{bfkl} -- the modification of the Bethe-Salpeter (BS) equation where external 
particles (quarks or gluons) and their propagators are replaced with the so-called ``reggeized'' partons, i.e., 
reggeons with quantum numbers of quarks and gluons obtained via solving the BS-equation 
with some perturbative kernel. Under usage of the BFKL approach the asymptotic behavior of some 
Regge trajectories was determined in the region of high negative values of the 
argument \cite{kwiecinski,kirlip} where perturbative 
techniques are applicable. However, the problem of calculating renorm-invariant Regge trajectories in the 
region of small values of the argument (or, at least, renorm-invariant intercepts of Regge trajectories) 
is so far unsolved.
The calculation of the ``hard pomeron'' (the leading reggeon associated with the spectrum of states 
formed by two ``reggeized'' gluons) intercept in the framework of the BFKL approach brings to the result 
explicitly depending on both the renormalization scheme and 
the renormalization scale \cite{camici}: 
$$
\alpha_{gg}(0) = 1+\frac{12\,\ln 2}{\pi}\alpha_s(\mu)
\left(1-\frac{20}{\pi}\alpha_s(\mu)\right)+o(\alpha^2_s(\mu))\,,
$$
where $\mu$ is the renormalization scale and $\alpha_s(\mu)=\frac{\bar g_s^2(\mu)}{4\pi}$ is the QCD effective 
running coupling. 

The direct aim of this paper is the calculation of renorm-invariant intercepts of Regge trajectories 
associated with colorless hadronic states formed by quark and antiquark. For this purpose 
we will use a technique alternative to the BFKL method that was applied earlier by C. Lovelace to 
the scalar field model $\phi^3_6$ \cite{lovelace}. This technique consists in looking for singularities of 
the solution of the BS-equation with some perturbative kernel and implies 
usage of effective running coupling. The basic physical idea follows. 
The Green function singularity points associated with the complex value 
of the quark-antiquark system invariant mass should not depend on the quark virtualities (this is the 
main hypothesis based on the fact that Regge trajectories are functions of one dynamical variable). So 
we may give virtualities of the quark and the antiquark asymptotically high negative values (but the value 
of the system invariant mass itself can be put low or even equal to zero). Considering the BS-equation 
in the region of asymptotically high quark virtualities allows us to choose the renormalization 
scale large enough for having a possibility to use the approximation of perturbative kernel minimal 
on powers of the running coupling (instead of the full renorm-invariant kernel). 
Such an approximation is called the ``ladder'' one. At this, the running 
coupling itself demonstrates characteristic perturbative behavior at the chosen renormalization scale 
(connected with high virtualities of the quarks). The positions of Green 
function singularities must depend on the quark-antiquark system invariant mass and QCD fundamental 
parameters only. Renorm-invariance should reveal itself in the fact that calculated Regge trajectories 
must be independent of the renormalization scheme.

\section*{The ladder approximation and intercepts of meson Regge trajectories}

In the case of zero invariant mass, $t=0$, the BS-equation for the full four-quark Green function 
$\hat G(q,p;\mu;\bar g_s(\mu))$ (here $q$ and $p$ are 4-momentums of quarks) cut at two points is of the form 

\vskip 1cm

\begin{picture}(460,40)

\ArrowLine(0,35)(40,35)
\ArrowLine(40,5)(0,5)
\Line(80,0)(80,40)
\Line(80,40)(40,40)
\Line(40,0)(40,40)
\Line(80,0)(40,0)
\Text(60,20)[c]{{\huge G}}
\Text(10,20)[c]{t=0}
\Text(20,40)[b]{{\large q}}
\Text(20,0)[t]{{\large q}}
\Text(90,40)[b]{{\large p}}
\Text(90,0)[t]{{\large p}}

\Text(110,20)[c]{{\Huge =}}
\Text(175,21)[c]{{\large $i(2\pi)^4\delta^4(q-p)$}}
\Text(240,21)[c]{{\Huge +}}

\ArrowLine(260,35)(300,35)
\ArrowLine(300,5)(260,5)
\Line(340,0)(340,40)
\Line(340,40)(300,40)
\Line(300,0)(300,40)
\Line(340,0)(300,0)
\ArrowLine(340,35)(380,35)
\ArrowLine(380,5)(340,5)
\Line(420,0)(420,40)
\Line(420,40)(380,40)
\Line(380,0)(380,40)
\Line(420,0)(380,0)
\Text(320,20)[c]{{\huge K}}
\Text(400,20)[c]{{\huge G}}
\Text(280,40)[b]{{\large q}}
\Text(280,0)[t]{{\large q}}
\Text(360,40)[b]{{\large k}}
\Text(360,0)[t]{{\large k}}
\Text(430,40)[b]{{\large p}}
\Text(430,0)[t]{{\large p}}

\end{picture}

\vskip 1cm

For the full renorm-invariant kernel $\hat K(q,p;\mu;\bar g_s(\mu))$ there takes place a formal expansion on 
powers of the effective running coupling: 

\vskip 1cm

\begin{picture}(300,40)

\ArrowLine(50,35)(90,35)
\ArrowLine(90,5)(50,5)
\Line(130,0)(130,40)
\Line(130,40)(90,40)
\Line(90,0)(90,40)
\Line(130,0)(90,0)
\Text(110,20)[c]{{\Huge K}}

\Text(160,20)[c]{{\Huge =}}

\ArrowLine(190,40)(230,40)
\ArrowLine(230,0)(190,0)
\Photon(230,0)(230,40){3}{5}
\Vertex(230,40){2}
\Vertex(230,0){2}
\Text(242,45)[c]{{\large $\bar g_s$}}
\Text(242,-5)[c]{{\large $\bar g_s$}}

\Text(280,20)[c]{{\Huge +}}

\Text(330,20)[c]{{\Large O($\bar g_s^4$).}}

\end{picture}

\vskip 1cm

If there is a hard dynamical scale in the BS-equation (for example, large virtuality of quarks), then by proper 
choice of the renormalization scale one can make terms of order O($\bar g_s^4$) in the kernel expansion 
to be little enough (in comparison with the tree term) to have a possibility to neglect them.
We are interested in those singularities which depend on the invariant mass $t$ of the system ``quark-antiquark'' 
and do not depend on other dynamical scales. So for consideration we may choose the region of large quark 
virtualities. At this, there emerges a possibility of usage of the minimal perturbative kernel and perturbative 
expressions for effective running parameters at scales of order of the quark virtualities.

In such an approximation the connected part of $\hat G$ turns out to be equal to the sum of the ladder diagrams:

\vskip 1cm

\begin{picture}(450,40)

\ArrowLine(0,40)(50,40)
\ArrowLine(50,0)(0,0)
\Photon(50,0)(50,40){3}{5}
\Vertex(50,40){2}
\Vertex(50,0){2}

\Text(75,20)[c]{{\large +}}

\ArrowLine(90,40)(140,40)
\ArrowLine(140,0)(90,0)
\Photon(140,0)(140,40){3}{5}
\Vertex(140,40){2}
\Vertex(140,0){2}
\ArrowLine(140,40)(190,40)
\ArrowLine(190,0)(140,0)
\Photon(190,0)(190,40){3}{5}
\Vertex(190,40){2}
\Vertex(190,0){2}

\Text(215,20)[c]{{\large +}}

\ArrowLine(240,40)(290,40)
\ArrowLine(290,0)(240,0)
\Photon(290,0)(290,40){3}{5}
\Vertex(290,40){2}
\Vertex(290,0){2}
\ArrowLine(290,40)(340,40)
\ArrowLine(340,0)(290,0)
\Photon(340,0)(340,40){3}{5}
\Vertex(340,40){2}
\Vertex(340,0){2}
\ArrowLine(340,40)(390,40)
\ArrowLine(390,0)(340,0)
\Photon(390,0)(390,40){3}{5}
\Vertex(390,40){2}
\Vertex(390,0){2}

\Text(435,20)[c]{{\large +$\;\;\;$. . .}}

\end{picture}

\vskip 1cm

Since we deal with the sum of the ladder diagrams (instead of the full Green function) the gauge and 
renorm-invariance of the results are not {\it a priori} (in contrast to the full kernel, the minimal perturbative 
kernel is not gauge and renorm-invariant) and should be grounded additionally. We will start from the 
arbitrary covariant gauge.

In this paper we restrict ourselves by looking for the colorless singularities at zero 
invariant mass, $t=0$ (intercepts of meson Regge trajectories), in the case of massless quarks. 
Upon the average over quark colors in the colorless $t$-channel, the BS-equation for $\hat G$ in 
the ladder approximation takes the form of (for better visual perception we omit spinor indices at 
matrix structures)
$$
\hat G(q,p)=i(2\pi)^4\delta^4(q-p)+
$$
$$
+\frac{N_c^2-1}{2N_c}\frac{\bar g_s^2(\sqrt{-q^2})}{i(2\pi)^4}\int \frac{d^4k}{(q-k)^2q^4}
\left(g_{\alpha\beta}-(1-\xi(\sqrt{-q^2}))\frac{(q-k)_\alpha(q-k)_\beta}{(q-k)^2}\right)
[\hat q\gamma^\alpha\hat G(k,p)\gamma^\beta\hat q]
$$
where $q$, $p$, and $k$ are quark 4-momentums, $\gamma^\alpha$ are Dirac matrices, 
$\hat q\equiv q_\mu\gamma^\mu$, $N_c$ is the number of colors, 
$\bar g_s(\mu)$ is the QCD effective running coupling, $\xi(\mu)$ is the running gauge parameter, and 
$\mu=\sqrt{-q^2}$ is the chosen renormalization scale 
(here and in the following, matrix structures to the left of the Green function should be attributed to the  
lower spinor line in the figures, and matrix structures to the right of $\hat G$ -- to the upper one).

Note that for large enough values of the renormalization scale the running gauge parameter 
behaves as 
$\xi(\mu)\sim \frac{1}{\ln\frac{\mu^2}{\Lambda^2}}\sim\bar g_s^2(\mu)$ \cite{yndurain}. 
So since we neglect loop corrections of order $\bar g_s^4$ and higher in the kernel, we may also neglect 
the longitudinal part of the gluon propagator in the tree term of the kernel because it also gives 
a contribution of order 
$\bar g_s^4$. In other words, in the region of asymptotically high $\sqrt{-q^2}$ 
the approximation of the minimal perturbative kernel (ladder approximation) does not 
contain dependence on the gauge.

After the convolution multiplication by $\hat q\times\hat q$ (the sign ``$\times$'' denotes the direct product) 
we come to the equation 
\begin{equation}
\label{bs}
q_\alpha q_\beta[\gamma^\alpha\hat G(q,p)\gamma^\beta]=i(2\pi)^4\delta^4(q-p)q_\alpha 
q_\beta[\gamma^\alpha\times\gamma^\beta]+
$$
$$
+\frac{N_c^2-1}{2N_c}\frac{\bar g_s^2(\sqrt{-q^2})}{i(2\pi)^4}\int \frac{d^4k}{(q-k)^2}
\left(g_{\alpha\beta}-\frac{(q-k)_\alpha(q-k)_\beta}{(q-k)^2}\right)[\gamma^\alpha\hat G(k,p)\gamma^\beta].
\end{equation}

Any ladder diagram can be represented as the sum of terms containing  
the direct product of two matrix structures. 
Since we put quarks massless, then each of these structures corresponds to 
the convolution product of the even number of 
Dirac matrices (vertices and quark propagators supply individual Dirac matrices
to the corresponding structure). So $\hat G(q,p)$ may be represented in form of 
$$
\hat G(q,p) = G^{00}(q, p)[I\times I] + 
\frac{1}{2}(G^{20}(q,p)(q_\mu p_\lambda - p_\mu q_\lambda) + 
E^{20}(q, p)e_{\mu\lambda\delta\eta}q^\delta p^\eta)[\sigma^{\mu\lambda}\times I] + 
$$
$$
+ \frac{1}{2}(G^{02}(q, p)(q_\nu p_\epsilon - p_\nu q_\epsilon) + 
F^{02}(q, p)e_{\nu\epsilon\xi\tau}q^\xi p^\tau)[I\times \sigma^{\nu\epsilon}] + 
$$
$$
+ \frac{1}{4}(G^{22}(q,p)(q_\mu p_\lambda - p_\mu q_\lambda)(q_\nu p_\epsilon - p_\nu q_\epsilon) + 
E^{22}(q,p)e_{\mu\lambda\delta\eta}q^\delta p^\eta(q_\nu p_\epsilon - p_\nu q_\epsilon) + 
$$
\begin{equation}
\label{struct}
+ F^{22}(q,p)(q_\mu p_\lambda - p_\mu q_\lambda)e_{\nu\epsilon\xi\tau}q^\xi p^\tau + 
H^{22}(q,p)e_{\mu\lambda\delta\eta}q^\delta p^\eta e_{\nu\epsilon\xi\tau}q^\xi p^\tau)
[\sigma^{\mu\lambda}\times \sigma^{\nu\epsilon}] + 
\end{equation}
$$
+ \frac{1}{2}(G^{24}(q,p)(q_\mu p_\lambda - p_\mu q_\lambda) + 
E^{24}(q, p)e_{\mu\lambda\delta\eta}q^\delta p^\eta)[\sigma^{\mu\lambda}\times \gamma^5] + 
$$
$$
+ \frac{1}{2}(G^{42}(q, p)(q_\nu p_\epsilon - p_\nu q_\epsilon) + 
F^{42}(q, p)e_{\nu\epsilon\xi\tau}q^\xi p^\tau)[\gamma^5\times \sigma^{\nu\epsilon}] + 
$$
$$
+ G^{40}(q, p)[\gamma^5\times I] + 
G^{04}(q, p)[I\times \gamma^5] + 
G^{44}(q, p)[\gamma^5\times \gamma^5].
$$
In square brackets there are Dirac matrix structures ($I$ is the unit matrix, 
$\sigma^{\mu\nu}\equiv i\frac{\gamma^\mu\gamma^\nu-\gamma^\nu\gamma^\mu}{2}$, 
$\gamma^5\equiv i\gamma^0\gamma^1\gamma^2\gamma^3$) and all unknown functions are scalars.

After the substitution of the last expression into (\ref{bs}) we convolve it with the  
structure $\frac{g_{\rho\sigma}}{16}[\gamma^\rho\times\gamma^\sigma+D^\rho\times D^\sigma]$ 
(here $D^\rho\equiv i\gamma^\rho\gamma^5$) over all spinor indices.

In view of relations 
$$
\frac{g_{\rho\sigma}}{16}\,q_\alpha q_\beta\,Sp[\gamma^\rho\gamma^\alpha\hat G(q,p)\gamma^\beta\gamma^\sigma + 
D^\rho\gamma^\alpha\hat G(q,p)\gamma^\beta D^\sigma] = q^2 \,\tilde G(q,p)
$$
and
$$
\frac{g_{\rho\sigma}}{16}\left(g_{\alpha\beta}-\frac{(q-k)_\alpha(q-k)_\beta}{(q-k)^2}\right)
Sp[\gamma^\rho\gamma^\alpha\hat G(q,p)\gamma^\beta\gamma^\sigma + 
D^\rho\gamma^\alpha\hat G(k,p)\gamma^\beta D^\sigma] = 3 \,\tilde G(k,p)\,,
$$
where $\tilde G(q,p)\equiv G^{00}(q, p)+(G^{22}(q, p)-H^{22}(q, p))(p^2q^2-(pq)^2)+G^{44}(q, p)$, 
we come to the scalar integral equation 
\begin{equation}
\label{bsscal}
q^2 \,\tilde G(q,p)=i(2\pi)^4\delta^4(q-p)\,q^2+
\frac{N_c^2-1}{2N_c}\frac{3\,\bar g_s^2(\sqrt{-q^2})}{i(2\pi)^4}\int \frac{d^4k}{(q-k)^2}\tilde G(k,p)
\end{equation}
for some linear combination of structure functions, $\tilde G(q,p)$. In other words, we have implemented a partial 
diagonalization of Eq. (\ref{bs}). 

If we put 
$\bar g_s^2(\sqrt{-q^2}) = \frac{16\pi^2}{\left(\frac{11}{3}N_c-\frac{2}{3}n_f\right)\ln\frac{-q^2}{\Lambda^2}}$ 
(where $n_f$ is the full number of quark flavors), 
Eq. (\ref{bsscal}) can be solved directly. 
In this way we can obtain some part of the spectrum of intercepts of meson Regge trajectories. The one-loop 
approximation of the running coupling $\bar g_s^2(\mu)$ 
is not renorm-invariant. It contains the QCD dimensional parameter $\Lambda$ depending on the renormalization scheme. 
Also, its value explicitly depends on the renormalization scale (above we have chosen 
$\mu=\sqrt{-q^2}$). 
But we are interested in those singularities of $\tilde G(q,p)$ which are renorm-invariant, i.e. 
independent of the renormalization scale (quark virtualities) and $\Lambda$. 
Only such singularities should be considered as intercepts of 
Regge trajectories.

The procedure of solving (\ref{bsscal}) is expounded in the Appendix. Here we give only the obtained 
series of intercepts 
\begin{equation}
\label{seria}
\alpha^{(k)}_{\bar q q}(0)=\frac{9(N_c^2-1)}{(2k+1)N_c(11N_c-2n_f)}-1\,,
\end{equation}
where $k$ is an arbitrary integer and $n_f$ is the full number of quark flavors. 

\section*{Discussion}

Let us point out the main features of series (\ref{seria}). The calculated intercepts depend on the fundamental 
dimensionless parameters, $N_c$ and $n_f$, and do not depend on quark virtualities, dimensional QCD 
parameter $\Lambda$, and the coupling. They do not depend on the renormalization scheme either.

The spectrum of intercepts from (\ref{seria}) will not change if we multiply the renormalization scale $\mu$ (fixed in the BS-equation) 
by arbitrary constant, i.e., if we put $\mu=C\sqrt{-q^2}$ instead of $\mu=\sqrt{-q^2}$. This is the straight consequence of 
their independence of $q^2$, $p^2$, and $\Lambda$. It is curious that independence of the coupling directly follows from the 
renorm-invariance of Regge trajectories in asymptotically free massless field models \cite{petrov}. 
All these facts point to renorm-invariance of series (\ref{seria}). 
The gauge invariance follows from the independence of the minimal perturbative kernel 
of the gauge parameter (this was discussed above). We would like to emphasize that the obtained 
result (for the case of massless quarks) is asymptotically accurate since the value of $-q^2$ is chosen to be asymptotically high.

In our calculations the masslessness of quarks (chiral symmetry) was an essential point. In reality the chiral symmetry is broken, 
either explicitly (in the Lagrangian) or dynamically, and the question how 
spectrum (\ref{seria}) is affected by nonzero quark masses is very important. Mass terms are present 
in both the quark propagators and the effective running coupling. However, even the heaviest quark mass does not bias 
the asymptotical behavior of the effective running coupling in the region of extra-high values of the renormalization scale. 
Relative to quark propagators the situation is more complicated. For massive quarks the BS-equation takes the form 
$$
(\hat q-m_1)\hat G(q,p)(\hat q-m_2)=i(2\pi)^4\delta^4(q-p)[(\hat q-m_1)\times(\hat q-m_2)]+\hat A(q,p)\,,
$$
where $m_1$ and $m_2$ are the quark masses and $\hat A(q,p)$ is the second (integral) term in the right-hand side of Eq. 
(\ref{bs}). In this case the numbers of both the Dirac matrix structures and the independent structure functions in (\ref{struct}) 
appreciably increase and partial diagonalization of the BS-equation cannot be implemented in such a simple way as we do for 
deriving (\ref{bsscal}) from (\ref{bs}). So at the moment we are not able to give quantitative estimation of the chiral symmetry breaking 
influence on series (\ref{seria}) explicitly from the BS-equation. But it is clear that increasing of quark masses should increase 
masses of meson states placed on Regge trajectories and, hence, decrease corresponding intercepts. Below we will see that 
comparison of (\ref{seria}) with meson phenomenology allows us to estimate this decrease for the leading ($k=0$) intercept from 
(\ref{seria}). Besides, such decreasing of intercepts implies that spectrum (\ref{seria}) can be associated only with 
those observed reggeon families which have all intercepts lower than the leading intercept from (\ref{seria}).

Let us consider light (i.e., composed of $u$- and $d$-quarks) mesons with isospin $I=1$. This choice is motivated 
by the wish to avoid mixing with more heavy quark flavors. For such mesons there exist four families of Regge 
trajectories -- $a$-reggeons, $\rho$-reggeons, $\pi$-reggeons, and $b$-reggeons. Chew-Frautschi plots (dependences of resonance 
spins on their masses squared) for these reggeons demonstrate unified linear behavior\footnote{This, of course, does not imply 
that true Regge trajectories are strictly linear since equations like $\alpha(M^2-{\rm i} M\Gamma)=N$ do not imply that 
${\rm Re}\,\alpha(M^2)=N$ and ${\rm Im}\,\alpha(M^2)=0$. Unfortunately, for most of the reggeons the only way to estimate their 
intercepts phenomenologically is the continuation of Chew-Frautschi plots.} with slopes about 1 GeV$^{-2}$ \cite{collins}. 
At $N_c=3$ and $n_f=6$ (three generations of fermions in the Standard Model) the leading intercept from (\ref{seria}) is 
$\alpha^{(0)}_{\bar q q}(0)=1/7$. Even-spin state reggeons with positive intercepts should not have a spin-0 state 
because it would have imaginary mass (tachyon). The leading $\pi$-trajectory has a spin-0 state, pion. Hence, 
if we assume the existence of the continious chiral limit, spectrum 
(\ref{seria}) cannot be associated with $\pi$-reggeons. The leading intercepts of $\rho$- and 
$a$-reggeons can be estimated not only via linear continuation of their Chew-Frautschi plots but also 
from the high energy evolution of angular distributions for reactions $\pi^-+p\to \pi^0+n$ and $\pi^-+p\to \eta+n$.
They both have values of about 0.45 \cite{irving}. Consequently, series (\ref{seria}) can be associated only with $b$-reggeons.
In Table \ref{tab1} we compare intercepts from series (\ref{seria}) at $N_c=3$ and $n_f=6$ with the observed states of $b$-reggeons.

\begin{table}[h]
\begin{center}
\begin{tabular}{|l|l|l|}
\hline
$k$   & $\alpha^{(k)}_{\bar q q}(0)$ & {$b\,$-mesons} \\
\hline
$0$   & $1/7$ & $b_1(1235)$ \\
            &           & $b_3(2025)$ \\
\hline
$1$   & $-13/21$ & {\bf $b_1(1960)$} \\
            &                & $b_3(2245)$ \\
\hline
$2$   & $-27/35$ & $b_1(2240)$ \\
\hline
\end{tabular}
\end{center}
\caption{The comparison of intercepts from (\ref{seria}) for the case $N_c=3$ and $n_f=6$ with the observed states of $b$-reggeons. 
The subscripts denote resonance spins and the numbers in parentheses denote approximate values of meson masses in MeV.}
\label{tab1}
\end{table}

\begin{figure}[ht]
\begin{center}
\vskip -3.2cm
\epsfxsize=7.5cm\epsfysize=7.5cm\epsffile{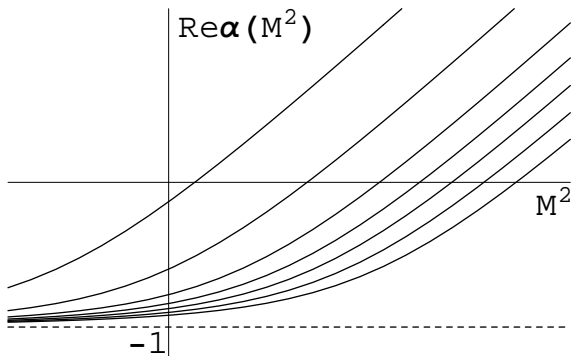}
\caption{Expected behavior of odd-spin meson Regge trajectories with negative intercepts.}
\label{bmes}
\end{center}
\end{figure}

Series (\ref{seria}) accumulates towards the value $\alpha^{(\infty)}_{\bar q q}(0)=-1$ (similar accumulations take place for 
the four-gluon Green function \cite{heckathorn}). However, the intercept of the Chew-Frautschi plot corresponding to the next 
to leading ($k=1$) $b$-trajectory is lower than $-1$. The same takes place for $\rho$-reggeons: Chew-Frautschi plots 
for all observed $\rho$-trajectories, except the leading one, have intercepts lower than $-1$ (the list of all observed 
$\rho$- and other mesons can be found at \cite{mesons}). But any true odd-spin Regge trajectory can take value $-1$ nowhere 
because this would cause a scattering amplitude unphysical singularity associated with some negative-spin state. Consequently, 
if we assume monotony of Regge trajectories below the first threshold, then all odd-spin Regge trajectories with 
negative intercepts should behave as in Fig.\ref{bmes}. Such a pattern seems quite natural. 
The nonlinearity of trajectories gains strength with radial quantum number $k$ increasing, and true values of reggeon 
intercepts should have values higher than those of their Chew-Frautschi plots. 

Now we can estimate the degree of influence of chiral symmetry breaking on the leading intercept value from series (\ref{seria}). 
The true intercept of the leading $b$-reggeon must have value lower than $\alpha^{(0)}_{\bar q q}(0)=1/7$ and higher than the 
intercept of the corresponding Chew-Frautschi plot, $\alpha_{\rm lin}^{(0)}=-0.17\pm 0.04$. For other $b$-reggeons we cannot do such a 
concrete estimation, but it is clear that they should have intercept values higher than $-1$ and the accumulation of these intercepts 
as in (\ref{seria}) is preserved at chiral symmetry breaking. So there takes place a qualitative correlation between 
the obtained series (\ref{seria}) and the phenomenology of $b$-mesons.

At the very end we must note that, in fact, we have not solved the BS-equation (\ref{bs}) completely and found only that part of 
the spectrum of intercepts of meson Regge trajectories which corresponds to $b$-reggeons. The full solution of Eq. (\ref{bs}) 
should also contain other parts of the four-quark Green function singularities which could be associated with 
$a$-, $\rho$-, and $\pi$-reggeons. Together with the problem of the chiral symmetry breaking influence on reggeon spectrums 
there is much further work to be done on these questions. 

\section*{Acknowledgment}

The author is indebted to V.A. Petrov for numerous and helpful discussions.

\section*{Appendix}

We will solve the integral equation 
\begin{equation}
\label{bsscalp}
q^2 \,\tilde G(q,p)=i(2\pi)^4\delta^4(q-p)\,q^2+
\frac{N_c^2-1}{2N_c}\frac{3\,\bar g_s^2(\sqrt{-q^2})}{i(2\pi)^4}\int \frac{d^4k}{(q-k)^2}\tilde G(k,p)
\end{equation}
using the technique applied earlier to the BS-equation in the scalar field model $\phi^3_6$ \cite{lovelace}. 

After the Wick rotation Eq. (\ref{bsscalp}) turns into 
\begin{equation}
\label{bsscalw}
\vec q\,^2 \tilde G(\vec q, \vec p)=(2\pi)^4\delta^4(\vec q-\vec p)\vec q\,^2+
\frac{N_c^2-1}{2N_c}\frac{3\,\bar g_s^2(|\vec q|)}{(2\pi)^4}\int \frac{d\vec k}{(\vec q-\vec k)^2}\tilde G(\vec k, \vec p)\,.
\end{equation}

Similarly to the three-dimensional case the amplitude $\tilde G$ can be expanded into a series of partial harmonics relative to the scattering angle 
$$
\cos\theta = \frac{(\vec p\, \vec q)}{|\vec p|\,|\vec q|}\,.
$$
A more detailed description of this formalism can be found in \cite{lovelace,erdelyi}. We restrict ourselves 
to the list of basic formulas.

In four space dimensions the partial wave expansion is of the form 
$$
T(\cos\theta)=\sum_{L=0}^\infty(2L+2)\,U_L(\cos\theta)\,T_L\,,
$$
where $U_L(x)$ are the second kind Chebyshev polynomials. The reverse formula is 
$$
T_L=\frac{1}{\pi(L+1)}\int_{-1}^1dx\sqrt{1-x^2}\,U_L(x)\,T(x)\,.
$$
If scalar functions $A$, $B$, and $C$ are related by 
$$
A(\vec q, \vec p)=\frac{1}{(2\pi)^4}\int d\vec k\,B(\vec q, \vec k)C(\vec k, \vec p)\,,
$$
then for the partial wave projections (in what follows, $q\equiv |\vec q|$, $k\equiv |\vec k|$, $p\equiv |\vec p|$)
$$
A_L(q, p)=\frac{1}{4\pi^2}\int_0^\infty dk\, k^3\, B_L(q, k)C_L(k, p)\,.
$$
And, at last, the definition of the four-dimensional analog of the second-kind Legendre function is 
$$
Q^{(4)}_L(z)=\frac{1}{\pi(L+1)}\int_{-1}^1dx\,\sqrt{1-x^2}\,\frac{U_L(x)}{z-x}\,.
$$
For this function the following integral representation exists:
$$
Q^{(4)}_L(\cosh \tau)=\int_{|\tau|}^\infty d\tau'e^{-(L+1)\tau'}=\frac{e^{-(L+1)|\tau|}}{L+1}\,.
$$

Now we introduce a dimensionless function $G_L(\ln q,\ln p) \equiv \frac{\tilde G_L(q, p)}{4\pi^2q^3p}$. 
Then, after proceeding in (\ref{bsscalw}) to the partial wave representation we come to 
\begin{equation}
\label{bsscalpar}
G_L(\ln q, \ln p)=\delta(\ln q-\ln p)+
\frac{N_c^2-1}{2N_c}\frac{3\,\bar g_s^2(q)}{8\pi^2}\int d (\ln k)\,Q^{(4)}_L\left(\cosh \left(\ln\frac{k}{q}\right)\right)\,G_L(\ln k, \ln p)\,.
\end{equation}

We define the Fourier transform of $G_L(\ln q, \ln p)$ as 
$$
G_L(\ln q, \ln p)\equiv \frac{1}{2\pi}\int_{-\infty}^\infty db\int_{-\infty}^\infty db'e^{i(b\ln q-b'\ln p)}[(L+1)^2+b^2]\tilde F_L(b,b')\,,
$$
where 
$$
F_L(\ln q, \ln p)\equiv \frac{1}{2\pi}\int_{-\infty}^\infty db\int_{-\infty}^\infty db'e^{i(b\ln q-b'\ln p)}\tilde F_L(b,b')\,. 
$$
Taking into account that 
$$
\int_{-\infty}^\infty d\tau e^{i\,b\,\tau}Q^{(4)}_L(\cosh \tau)=\frac{2}{(L+1)^2+b^2}
$$
we come to the equation 
$$
\frac{1}{2\pi}\int_{-\infty}^\infty db\int_{-\infty}^\infty db'e^{i(b\ln q-b'\ln p)}[(L+1)^2+b^2]\tilde F_L(b,b')=\delta(\ln q-\ln p)+
$$
$$
+\frac{N_c^2-1}{2N_c}\frac{3\,\bar g_s^2(q)}{4\pi^2}\frac{1}{2\pi}\int_{-\infty}^\infty db\int_{-\infty}^\infty db'e^{i(b\ln q-b'\ln p)}\tilde F_L(b,b')
$$
which is, in turn, equivalent to the differential equation 
\begin{equation}
\label{differ}
\left[(L+1)^2-\frac{d^2}{d(\ln q)^2}\right]F_L(\ln q, \ln p)=\delta(\ln q-\ln p)+\frac{N_c^2-1}{2N_c}\frac{3\,\bar g_s^2(q)}{4\pi^2}F_L(\ln q, \ln p)\,.
\end{equation}
In what follows, at asymptotically high values of $q$ we will use the perturbative one-loop approximation to the 
QCD effective running coupling, 
$\bar g_s^2(q) = \frac{8\pi^2}{\left(\frac{11}{3}N_c-\frac{2}{3}n_f\right)\ln\frac{q}{\Lambda}}$ (at this, 
all the dependence on the renormalization scheme is in the constant $\Lambda$).   
So the Fourier transform of (\ref{differ}) takes the form of 
\begin{equation}
\label{integr}
[(L+1)^2+b^2]\tilde F_L(b,b')=\delta(b-b')-\frac{i\,E}{2}\int_{-\infty}^\infty db''\epsilon(b-b'')\tilde F_L(b'',b')\,,
\end{equation}
where $\epsilon(b)=b/|b|$ and $E\equiv \frac{9(N_c^2-1)}{N_c(11N_c-2n_f)}$. 

Now it is convenient to introduce a new variable, 
$$
\chi\equiv\int_0^b\frac{db''}{(L+1)^2+b''^2}=\frac{\arctan\frac{b}{L+1}}{L+1}\,,\;\;\;\;\Omega\equiv\chi(\infty)=\frac{\pi}{2(L+1)}.
$$

Function $\chi(b)$ maps $b=(-\infty,\infty)$ onto $\chi=(-\Omega,\Omega)$ monotonically. So 
$$
\epsilon(b-b')=\epsilon(\chi-\chi')\,,\;\;\;\;\delta(b-b')((L+1)^2+b'^2)=\delta(\chi-\chi')\,,
$$
where $\chi'\equiv \chi(b')$ and in terms of the new function 
$y_L(\chi,\chi')\equiv ((L+1)^2+b^2)\tilde F_L(b,b')((L+1)^2+b'^2)$ Eq. (\ref{integr}) takes the form of 
$$
y_L(\chi,\chi')=\delta(\chi-\chi')-\frac{i\,E}{2}\int_{-\infty}^\infty d\chi''\epsilon(\chi-\chi'')\tilde y_L(\chi'',\chi')\,.
$$
After substitution $y_L(\chi,\chi')=\frac{\partial j(\chi,\chi')}{\partial \chi}$ we come to differential equation  
$$
\delta(\chi-\chi')+\frac{i\,E}{2}\{j(\Omega,\chi')+j(-\Omega,\chi')\}=
e^{-i\,E\,\chi}\frac{\partial}{\partial \chi}\left\{e^{i\,E\,\chi}j(\chi,\chi')\right\}\,.
$$
The solution of this equation is 
$$
e^{i\,E\,\chi}j(\chi,\chi')=e^{i\,E\,\chi'}\left[C+\frac{1}{2}\epsilon(\chi-\chi')\right]+
\frac{1}{2}e^{i\,E\,\chi}\{j(\Omega,\chi')+j(-\Omega,\chi')\}\,.
$$
The value of $C$ is not arbitrary. 
Since
$$
j(\pm\Omega,\chi')=e^{i\,E(\chi'\mp\Omega)}\left[C\pm\frac{1}{2}\right]+
\frac{1}{2}\{j(\Omega,\chi')+j(-\Omega,\chi')\}\,,
$$
then adding $\pm$ forms of this equation, we find $C=\frac{i}{2}\tan(E\Omega)$.
Hence, finally, the solution of (\ref{integr}) is of the form 
$$
[(L+1)^2+b^2]\tilde F_L(b,b')=\delta(b-b')-\frac{i\,E}{2}e^{i\,E\,(\chi(b')-\chi(b))}[\epsilon(\chi-\chi')+i\tan(E\Omega)]
$$
from where we obtain a series of the amplitude singularities (intercepts of meson Regge trajectories)
$$
\alpha^{(k)}_{\bar q q}(0)=\frac{9(N_c^2-1)}{(2k+1)N_c(11N_c-2n_f)}-1\,,
$$
where $k$ is an arbitrary integer.


\begin{thebibliography}{99}

\bibitem{collins} P.D.B. Collins, {\bf An Introduction to Regge Theory  
                  $\&$ High Energy Physics}.\\ Cambridge University Press 1977

\bibitem{bfkl} E.A. Kuraev, L.N. Lipatov, V.S. Fadin,
               Sov.Phys.JETP {\bf 44} (1976) 443\\
               Ya.Ya. Balitsky, L.N. Lipatov, 
               Sov.J.Nucl.Phys. {\bf 28} (1978) 822

\bibitem{kwiecinski} J. Kwiecinski, Phys.Rev. D {\bf 26} (1982) 3293
\bibitem{kirlip} R. Kirschner, L.N. Lipatov, Z.Phys. C {\bf 45} (1990) 477

\bibitem{camici} V.S. Fadin, L.N. Lipatov, Phys.Lett. B {\bf 429} (1998) 127\\
                 G. Camici, M. Ciafaloni, Phys.Lett. B {\bf 430} (1998) 349

\bibitem{lovelace} C. Lovelace, Nucl.Phys. B {\bf 95} (1975) 12

\bibitem{yndurain} F.J. Yndurain, {\bf Quantum Chromodynamics}. 
                   Springer-Verlag 1983\\
                   S. Narison, Phys.Rept. {\bf 84} (1982) 263

\bibitem{petrov} V.A. Petrov, hep-ph/0603103

\bibitem{irving} A.V. Barnes et al., Phys.Rev.Lett. {\bf 37} (1976) 76\\
                 O.I. Dahl et al., Phys.Rev.Lett. {\bf 37} (1976) 80\\
                 A.C. Irving, R.P. Worden, Phys.Rept. {\bf 34} (1977) 117

\bibitem{heckathorn} D. Heckathorn, Phys.Rev. D {\bf 18} (1978) 1286

\bibitem{mesons} Particle Data Group, http://pdg.lbl.gov/2009/mcdata/mass\_width\_2008.csv

\bibitem{erdelyi} A. Erdelyi et al., {\bf Higher Transcendental Functions}. 
                  McGraw-Hill 1953

\end{thebibliography}
\end{document}